\documentclass[fleqn,usenatbib,useAMS]{mnras}

\usepackage{newtxtext,newtxmath}

\usepackage[T1]{fontenc}
\usepackage{ae,aecompl}
\usepackage{hyperref}

\usepackage{natbib}
\bibpunct{(}{)}{;}{a}{,}{,}

\usepackage{graphicx}	
\usepackage{amsmath}	
\usepackage{amssymb}	

\usepackage{threeparttable}
\usepackage{lipsum}
\usepackage{mathtools}
\usepackage{cuted}
\usepackage{float}

\title[Blazar dichotomy as magnetized accretion process]{FSRQ/BL Lac dichotomy as the magnetized advective accretion process around black holes: a unified classification of blazars}

\author[T. Mondal and B. Mukhopadhyay]{
Tushar Mondal\thanks{mtushar@iisc.ac.in (TM)} and
Banibrata Mukhopadhyay\thanks{bm@iisc.ac.in (BM)}
\\
Department of Physics, Indian Institute of Science, Bangalore 560012, India
}

\date{Accepted 2019 April 6. Received 2019 April 1}

\pubyear{2019}

\begin{document}
\label{firstpage}
\pagerange{\pageref{firstpage}--\pageref{lastpage}}
\maketitle
\begin{abstract}
The \textit{Fermi} blazar observations show a strong correlation between $\gamma$-ray luminosities and spectral indices. BL Lac objects are less luminous with harder spectra than flat-spectrum radio quasars (FSRQs). Interestingly FSRQs are evident to exhibit a Keplerian disc component along with a powerful jet. We compute the jet intrinsic luminosities by beaming corrections determined by different cooling mechanisms. Observed $\gamma$-ray luminosities and spectroscopic measurements of broad emission lines suggest a correlation of the accretion disc luminosity with jet intrinsic luminosity. Also, theoretical and observational inferences for these jetted sources indicate a signature of hot advective accretion flow and a dynamically dominant magnetic field at jet-footprint. Indeed it is difficult to imagine the powerful jet launching from a geometrically thin Keplerian disc. We propose a  magnetized, advective disc-outflow symbiosis with explicit cooling to address a unified classification of blazars by controlling both the mass accretion rate and magnetic field strength. The large scale strong magnetic fields influence the accretion dynamics, remove angular momentum from the infalling matter, help in the formation of strong outflows/jets, and lead to synchrotron emissions simultaneously. We suggest that the BL Lacs are more optically thin and magnetically dominated than FSRQs at the jet-footprint to explain their intrinsic $\gamma$-ray luminosities.	
\end{abstract}




\section{Introduction} \label{sec:intro}

Relativistic, highly collimated, powerful jets are ubiquitous in various astrophysical black hole (BH) systems spanning BHs from stellar mass $(\sim 10 \ \text{to}\ 30M_{\odot},\ M_{\odot}\ \text{is solar mass})$ to supermassive scales $(\sim 10^{6} \ \text{to}\ 10^{9}M_{\odot})$. They particularly exist in microquasars, active galactic nuclei (AGNs), and gamma-ray bursts (GRBs). These jets are believed to be produced from the innermost region of accretion flow, presumably when the accretion flow is hot, advective and geometrically thick. Observational evidences predict that every BH X-ray binary (XRB) in their low/hard state has strong radio jet \citep{2004MNRAS.355.1105F} and a parallel discovery is that all low luminous AGNs are radio-loud \citep{2000ApJ...542..186N}. There are indeed some causal connections between hot advective accretion flows and radio-emitting jets. Compared with cooler geometrically thin accretion discs \citep{1973A&A....24..337S}, hot advective flows can easily advect the magnetic field in the vicinity of BH \citep{2012MNRAS.424.2097G}. Also, the strong winds from the hot advective flows help to collimate \citep{2000AstL...26..208B} and stabilize the jet \citep{1992A&A...256..354A}.

Observations and theoretical efforts over decades try to address the mechanisms responsible for these jet formation, as well as the nature of their high-energy radiation \citep{2001Sci...291...84M}. 
The leading mechanism suggests that the combined effects of large-scale magnetic fields and BH rotation could produce efficient outflowing energy in jets \citep{1977MNRAS.179..433B}. This was also verified by three-dimensional, general relativistic, magnetohydrodynamic (GRMHD) simulations 
\citep{2004ApJ...611..977M,2011MNRAS.418L..79T,2018ApJ...859...28Q}. Indeed the inferred unusually large Faraday rotation from the multi-frequency radio measurement of a pulsar close to the center of our Galaxy indicates a signature of the dynamically important magnetic field near the BH \citep{2013Natur.501..391E}. Also based on the correlation of accretion disc luminosity and jet magnetic field for a sample of 76 radio-loud active galaxies, it was concluded that the jet-launching regions are threaded by dynamically important magnetic fields \citep{2014Natur.510..126Z}. All these inferences support the idea of highly magnetized accretion flow in the vicinity of a BH.

Now the question is, how are the jet power and the accretion disc luminosity correlated?   The intrinsic jet luminosity (the power that jet expends in producing non-thermal radiation) is believed to be $10\%$ of the jet kinetic power, and interestingly, this holds for all the well-known jetted sources, namely, BH XRBs, AGNs, and GRBs \citep{2012Sci...338.1445N, 2014ApJ...780L..14M}. Recently a clear correlation between the intrinsic jet luminosity, as measured through the $\gamma$-ray luminosity, and the accretion luminosity, as measured by the broad emission lines, for a sample of 217 blazar sources is found. Remarkably, both of them are of the same order \citep{2014Natur.515..376G}. Following all these observational and theoretical insights, we construct a magnetized disc-outflow symbiosis to address a unified classification of blazar sequences by controlling both the magnetic field strength and accretion rate.

Blazars are radio-loud AGNs with relativistic jets pointing close to our line of sight. Based on the equivalent width (EW) of the optical emission lines, blazars are classified into two subclasses: Flat Spectrum Radio Quasars (FSRQs) with EW$\ge 5$ {\AA} and BL Lac objects with EW$< 5$ {\AA} \citep{1995PASP..107..803U}. The signature of strong emission lines stipulates the presence of luminous broad line region (BLR) and, hence, efficient accretion process in FSRQs. On the other hand, the presence/absence of narrow emission lines in BL Lacs suggests a relatively low accretion and/or the ascendancy of non-thermal synchrotron emission
therein.

The broad-band spectral energy distribution (SED) of blazars consists of two broad humps. The low energy component extends from radio to optical-UV or X-rays and is well explained with synchrotron emission by relativistic electrons. The high energy component, generally in the $\gamma$-ray regime, is attributed to the inverse Compton (IC) scattering of soft photons by energetic electrons in jet plasma. In BL Lac objects, the soft photons are thought to be provided by the synchrotron emission and this IC mechanism is known as synchrotron self-Compton (SSC). On the other hand, for powerful FSRQs, the IC mechanism is explained by the combination of SSC and (mostly) external Comptonization (EC) processes. For EC, the soft photons are expected to be provided by the jet environment, either the accretion disc \citep{1993ApJ...416..458D} and/or BLR \citep{2009ApJ...704...38S} and/or dusty torus \citep{2000ApJ...545..107B}. Also, based on the position of the synchrotron peak $(\nu _{syn})$ in the rest frame, BL Lac objects and blazars in general are further classified as low-synchrotron-peaked (LSP, $\nu _{syn}<10^{14}$ Hz), intermediate-synchrotron-peaked (ISP, $10^{14} \ \text{Hz} <\nu _{syn}<10^{15}$ Hz), and high-synchrotron-peaked (HSP, $\nu _{syn}>10^{15}$ Hz) by \citet{2010ApJ...716...30A}.

In this paper, we infer the role of large-scale magnetic field in underlying dynamics and efficient cooling mechanisms for different blazar classes. The ordered magnetic field anchored in the accretion flow provides a very efficient way of tapping the gravitational potential energy of a BH and this energy is liberated in the form of outflows/jets \citep{1982MNRAS.199..883B}. Also the vertically inflated strong toroidal field, produced by the differential rotation of the accretion flow, can enhance the outflow power in the formation of `magnetic tower jets' \citep{2003MNRAS.341.1360L}. We suggest, based on our unified model, that BL Lac objects are more magnetized, advective and optically thin compared to the FSRQs at the jet-launching region to explain their intrinsic $\gamma$-ray luminosities. We aim at uncovering the physical reason behind the luminosity variation of different blazar classes and why softer FSRQs still exhibiting most powerful jets.

In next Section, we model the coupled disc-outflow system for magnetized advective accretion flows considering different radiation mechanisms. In Section \ref{sec:results}, we discuss the consequence of such flows, in particular focusing on the energetics of magnetized accretion flows to explain blazars, as well as different beaming statistics. Finally we end with discussions and conclusions in Sections~\ref{sec:discussions} and \ref{sec:conclusions} respectively.

\section{MODELING THE COUPLED MAGNETIZED DISK-OUTFLOW SYSTEM} \label{sec:model}

\subsection{General equations for magnetized advective accretion flow}
We propose a magnetized, viscous, advective disc-outflow/jet symbiotic system 
with cooling explicitly around BHs. 
Unlike previous exploration \cite[e.g.][]{2004ApJ...616..669K}, we consider 
the large scale magnetic and turbulent viscous stresses both.
Here we assume a steady and axisymmetric flow, hence $\partial / \partial t \equiv \partial / \partial \phi \equiv 0$ and that all the flow parameters: radial velocity $(v_{r})$, specific angular momentum $(\lambda)$, outflow or vertical velocity $(v_{z})$, fluid pressure $(p)$, mass density $(\rho)$, radial $(B_{r})$, azimuthal $(B_{\phi})$, and vertical $(B_{z})$ components of magnetic field, are functions of both radial and vertical coordinates. Throughout we express length variables in units of $GM_{BH}/c^{2}$, where $G$ is the Newton's gravitation constant, $M_{BH}$ the mass of BH, and $c$ the speed of light. Accordingly, we also express 
other variables. 
We make a reasonable hypothesis in the disc-outflow symbiotic region that, the vertical variations of any dynamical variables (say, $A$) is much less than that with radial variation, that allows us to introduce $\partial A/\partial z\approx sA/z$, where $s$ is just the degree of scaling which is a small number. These constants for dynamical variables $v_{r},\ \lambda,\ v_{z},\ B_{r},\ B_{\phi},\ B_{z},\ p,\ \text{and} \ \rho$ are $s_{1},\ s_{2},\ s_{3},\ s_{4},\ s_{5},\ s_{6},\ s_{7}, \ \text{and} \ s_{8}$ respectively. Hence, the continuity equation, the components for momentum balance equation, the equation for no magnetic monopole, and the components for induction equation are respectively
\begin{equation}
\frac{1}{r} \frac{\partial}{\partial r}\left(r\rho v_{r}\right)+\left(s_{3}+s_{8}\right)\frac{\rho v_{z}}{z}=0, \label{continuity}
\end{equation}
\begin{multline}
v_{r}\frac{\partial v_{r}}{\partial r}+s_{1}\frac{v_{r}v_{z}}{z}-\frac{\lambda^{2}}{r^{3}}+\frac{1}{\rho}\frac{\partial p}{\partial r}+F \\ 
=\frac{1}{4 \pi  \rho}\left[-\frac{B_{\phi}}{r}\frac{\partial}{\partial r}\left(rB_{\phi}\right)+B_{z}\left(s_{4}\frac{B_{r}}{z}-\frac{\partial B_{z}}{\partial r}\right)\right],
\end{multline}
\begin{multline}
v_{r}\frac{\partial \lambda}{\partial r}+s_{2}\frac{v_{z}\lambda}{z}=\Bigg[\frac{3\alpha}{\rho}\left(p+\rho v_{r}^{2}\right)+\frac{\alpha r}{\rho}\left(\frac{\partial p}{\partial r}+2\rho v_{r}\frac{\partial v_{r}}{\partial r}+v_{r}^{2}\frac{\partial \rho}{\partial r}\right) \\ 
+\frac{\alpha z}{\rho}\left(\frac{s_{7} p}{z}+2\rho v_{r}\frac{s_{1}v_{r}}{z}+v_{r}^{2}\frac{s_{8}\rho}{z}\right)\Bigg] \\ 
+\frac{r}{4\pi \rho}\left[\frac{B_{r}}{r}\frac{\partial}{\partial r}\left(rB_{\phi}\right)+s_{5}\frac{B_{z} B_{\phi}}{z}\right],
\end{multline}
\begin{multline}
v_{r}\frac{\partial v_{z}}{\partial r}+s_{3}\frac{v_{z}^{2}}{z}+s_{7}\frac{p}{\rho z}+\frac{z}{r}F= \\ 
\frac{1}{4 \pi  \rho}\left[B_{r}\left(\frac{\partial B_{z}}{\partial r}-s_{4}\frac{B_{r}}{z}\right)-s_{5}\frac{B_{\phi}^{2}}{z}\right],
\end{multline}
\begin{equation}
\frac{1}{r}\frac{\partial}{\partial r}\left(rB_{r}\right)+s_{6}\frac{B_{z}}{z}=0,
\end{equation}
\begin{equation}
\left(s_{3}+s_{4}\right)v_{z}B_{r}-\left(s_{1}+s_{6}\right)v_{r}B_{z}=0,
\end{equation}
\begin{equation}
\frac{\partial}{\partial r}\left(v_{r}B_{\phi}-\frac{\lambda B_{r}}{r}\right)-\left(s_{2}+s_{6}\right)\frac{\lambda B_{z}}{rz}+\left(s_{3}+s_{5}\right)\frac{v_{z} B_{\phi}}{z}=0,
\end{equation}
\begin{equation}
\frac{\partial}{\partial r}\left[r\left(v_{z}B_{r}-v_{r}B_{z}\right)\right]=0,
\end{equation}
where $F$ is the magnitude of the gravitational force for a BH in the pseudo-Newtonian framework \citep{2002ApJ...581..427M}. The set of equations without imposing vertical scaling was presented in \cite{2019MNRAS.482L..24M}. The importance of generalized viscous shearing stress tensor $(W_{ij})$ is taking care explicitly in the present formalism. Various components of $W_{ij}$ are written in terms of \citet{1973A&A....24..337S} $\alpha$-prescription with appropriate modifications \citep{2019MNRAS.482L..24M}.
We consider the induction equation in the limit of very large magnetic Reynolds number ($\propto 1/\nu_m$, where $\nu_{m}$ is the magnetic diffusivity), which is the case for an accretion disc. Note that at very large field, $W_{ij}$ is not expected to contribute, as they could govern by turbulence induced due to weak magnetic fields. In that situation, the accretion is solely due to large scale magnetic shear.

\subsection{Thermodynamics of the gas: effects of magnetic and radiation pressures}

In the presence of magnetic field, the standard equation of state for a mixture of perfect gas and radiation is
\begin{equation}
	p_{t}=p+p_{m}=p_{g}+p_{r}+p_{m}=\frac{\rho k_{B}T}{\mu m_{p}}+\frac{1}{3}aT^{4}+\frac{B^{2}}{8\pi},
\end{equation}
where $p_{t}$ is the total pressure, $k_{B}$ the Boltzmann constant, $\mu$ the mean molecular weight, $m_{p}$ the proton mass and $a$ the Stefan constant. $p_{g}=\beta p$ is the gas pressure, $p_{r}=(1-\beta)p$ is the radiation pressure, and $p_{m}=(\beta /\beta_{m})p$ is the magnetic pressure. Here we take the parameter $\beta = p_{g}/(p_{g}+p_{r})$ to be independent of $r$, unlike the parameter plasma-$\beta$, $\beta_{m}=(p_{g}/p_{m})$.

In the context of a frozen-in field, the internal energy per unit mass of the system is
\begin{equation}
	U=\frac{3}{2}\frac{\rho k_{B}T}{\mu m_{p}}V+aT^{4}V+\frac{B^{2}}{4\pi}V,
\end{equation}
where $V$ is the volume of unit mass of gas. From first law of thermodynamics and using flux-freezing assumption, the entropy gradient can be written in terms of temperature and density gradients as
\begin{equation}
	Tds=\frac{p}{\rho}\left[ \left( 12-\frac{21}{2}\beta \right) \frac{dT}{T}-\left(4-3\beta+\frac{1}{3}\frac{\beta}{\beta_{m}} \right) \frac{d\rho}{\rho} \right]. 
\end{equation}

\subsection{Radiation mechanisms for two temperature plasma}
We solve the energy balance equations for ions and electrons by taking into account the detailed balance of heating, cooling and advection. Since ions are much heavier than the electrons, we expect all the generated heats primarily act on the ions. Some part of this energy is transferred from ions to electrons through different thermal and non-thermal coupling mechanisms. The magnetized energy equation for ions reads as
\begin{equation}
\Gamma_{3} ' \left[ v_{r}\left\lbrace\frac{\partial p}{\partial r}-\Gamma_{1}\frac{p}{\rho}\frac{\partial \rho}{\partial r}\right\rbrace+v_{z}\left\lbrace\frac{\partial p}{\partial z}-\Gamma_{1}\frac{p}{\rho}\frac{\partial \rho}{\partial z}\right\rbrace \right] 
=Q^{+}-Q^{ie}, 
\label{eq:energy1}
\end{equation}
where
\begin{equation*}
	\Gamma_{1}=\frac{32-24\beta -3\beta^{2}+\frac{2\beta (4-3\beta)}{3\beta_{m}}}{24-21\beta} \ , \ \text{and} \ \Gamma_{3} '=\frac{24-21\beta}{2(4-3\beta)}.
\end{equation*}
The detailed description of the rate of energy generated per unit volume $(Q^{+})$ via magnetic and viscous dissipations is followed from our earlier work \citep{2019MNRAS.482L..24M}. If ions are at higher temperature than electrons, then the volume transfer rate of the energy from ions to electrons through Coulomb coupling is $Q^{ie}$ \citep{2010MNRAS.402..961R}. The electron energy equation is then given by
\begin{equation}
\Gamma_{3} ' \left[ v_{r}\left\lbrace\frac{\partial p_{e}}{\partial r}-\Gamma_{1}\frac{p_{e}}{\rho}\frac{\partial \rho}{\partial r}\right\rbrace+v_{z}\left\lbrace\frac{\partial p_{e}}{\partial z}-\Gamma_{1}\frac{p_{e}}{\rho}\frac{\partial \rho}{\partial z}\right\rbrace \right] 
=Q^{ie}-Q^{-}, 
\label{eq:energy2}
\end{equation}
where $Q^{-}$ represents the radiative cooling rate through electrons via different cooling processes including bremsstrahlung $(q_{br}^{-})$, synchrotron $(q_{syn}^{-})$ and inverse comptonization process off soft synchrotron photons $(q_{syn,C}^{-})$ and external soft photons coming mainly from disc $(q_{disc,C}^{-})$. The dimensionful radiative cooling rate per unit volume is
\begin{equation*}
q^{-}=Q^{-}c^{11}/(G^{4}M_{BH}^{3})=q_{br}^{-}+q_{syn}^{-}+q_{syn,C}^{-}+q_{disc,C}^{-}.
\end{equation*}
Various cooling formalisms are adopted from \citet{1995ApJ...452..710N} and \citet{2010MNRAS.402..961R}, where
\begin{multline}
q_{br}^{-}=1.4\times 10^{-27}n_{i}n_{e}T_{e}^{1/2}(1+4.4\times 10^{-10}T_{e}) \\ \text{ergs\ cm$^{-3}$\ s$^{-1}$},
\end{multline}
\begin{equation}
q_{syn}^{-}=\frac{2\pi}{3c^{2}}k_{B}T_{e}\frac{\nu _{c}^{3}}{R} \ \text{ergs\ cm$^{-3}$\ s$^{-1}$},
\end{equation}
\begin{equation}
q_{syn,C}^{-}=q_{syn}^{-}\left[\eta _{1}-\eta _{2}\left(\frac{x_{c}}{3\theta _{e}} \right)^{\eta _{3}} \right] \ \text{ergs\ cm$^{-3}$\ s$^{-1}$},
\end{equation}
\begin{multline}
q_{disc,C}^{-}=3\frac{F_{disc}}{R}\left(\frac{\theta _{e}}{x_{b}} \right)^{3}\Bigg[\frac{\eta_{1}}{3} \left\lbrace \left(\frac{x_{max}}{\theta_{e}} \right)^{3}-\left(\frac{x_{c}}{\theta_{e}} \right)^{3} \right\rbrace \\
 -\frac{\eta_{2}}{3+\eta_{3}} \left\lbrace \left(\frac{x_{max}}{\theta_{e}} \right)^{3+\eta_{3}}-\left(\frac{x_{c}}{\theta_{e}} \right)^{3+\eta_{3}} \right\rbrace \Bigg] \ \text{ergs\ cm$^{-3}$\ s$^{-1}$},
\end{multline}
and the synchrotron self-absorption cut-off frequency $(\nu_{c})$ is computed numerically at every radius $R \ (=rGM_{BH}/c^{2})$ following \cite{1995ApJ...452..710N}. From charge neutrality, the number density for ions $(n_{i})$ and electrons $(n_{e})$ are equal. The Comptonized energy enhancement factor $\eta$ is defined to be the average change in energy of a photon between injection and escape, and it is prescribed as \citep{1991ApJ...369..410D}
\begin{multline}
\eta = 1+\frac{P(A-1)}{1-PA}\left[1-\left(\frac{x}{3\theta_{e}}\right)^{-1-\ln P/\ln A} \right] \\
\equiv  1+\eta_{1}-\eta_{2}\left(\frac{x}{\theta_{e}}\right)^{\eta_{3}},
\end{multline}
where
\begin{multline*}
x=\frac{h\nu}{m_{e}c^{2}},\
\theta_{e}=\frac{k_{B}T_{e}}{m_{e}c^{2}}, \
P=1-exp(-\tau_{es}),\\
\tau_{es}=\kappa_{es}\rho H, \ 
A=1+4\theta_{e}+16\theta_{e}^{2}.
\end{multline*}
Here $P$ is the probability that an escaping photon is scattered, $\tau_{es}$ the scattering optical depth with $\kappa_{es}=0.38 \ \text{cm$^{2}$ g$^{-1}$}$, $T_{e}$ the electron temperature, $H$ half scale height and $A$ the mean amplification factor in the energy of a scattered photon when the scattering electrons have a Maxwellian velocity distribution of temperature $\theta_{e}$. For external comptonization of disc soft photons, the disc blackbody temperature $(T_{b})$ and disc flux $(F_{disc})$ are adopted from \cite{1973blho.conf..343N}. Here, the Rayleigh-Jeans $\nu^{2}$ spectrum cut-off frequency $\nu_{b}=5.61\times 10^{10}\ T_{b}$  is used to calculate the factors $x_{b}=h\nu_{b}/(m_{e}c^{2})$ and $x_{max}=\text{max}\ (x_{b},3\theta_{e})$.

\subsection{Method}

Very far away from the BH, the matter is sub-sonic, i.e. $v_{r}$ is very less than medium sound speed $c_{s}$. Near the event horizon, matter is supersonic. In between on the way, matter passes through the sonic/critical point, where $v_{r}\simeq c_{s}$. We use this critical point as one of the boundaries to solve the model coupled differential equations described in \S\S 2.1, 2.2, 2.3. The other two boundaries are as follow. Far away from the BH, the transition radius from the Keplerian to sub-Keplerian flows provides the outer boundary of our solution. The event horizon where the matter velocity reaches the light speed provides the inner boundary. To capture the sonic/critical point conditions, we combine the above fundamental model equations appropriately (see, e.g., \citealt{2010MNRAS.402..961R,2018MNRAS.476.2396M},
for detailed description). At the critical radius $r_{c}$, we need to specify the electron temperature $T_{ec}$ and specific angular momentum $\lambda_{c}$. For a realistic accretion, flow must exhibit at least one `saddle'-type critical point. We solve the above mentioned model magnetohydrodynamic equations including ion energy equation, further supplemented by the electron energy equation. Note that one needs to adjust self-consistently $r_{c},\ \lambda_{c},\ v_{zc},\ T_{ec}$ and relative dependence of magnetic field components in order to obtain the solutions connecting the outer boundary to the BH event horizon through $r_{c}$.

\section{Blazars as magnetized accretion-outflow/jet systems} \label{sec:results}

\subsection{Observation} \label{sec:observations}
\begin{figure*}
	\center
	\includegraphics[width=12 cm]{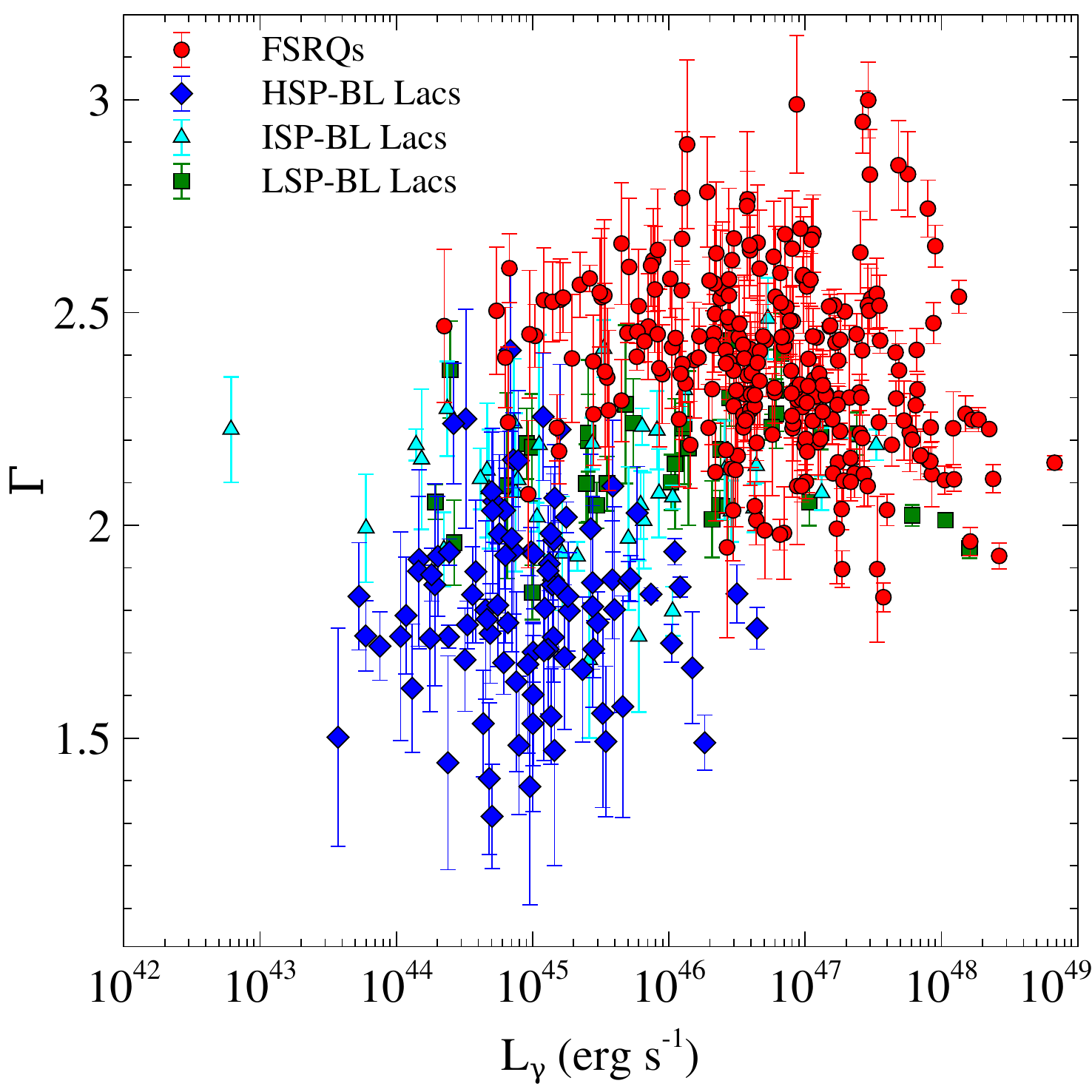}\caption{Photon spectral index versus $\gamma$-ray luminosity for FSRQs and different classes of BL Lac objects. 
	} 
	\label{fig:source}
\end{figure*}

We consider blazar samples detected in the $\gamma$-ray band by \textit{Fermi} LAT during its second catalog of AGNs \citep{2011ApJ...743..171A}. 
The sample contains 381 FSRQs and 440 BL Lacs. However, we consider only those sources for which redshift was measured. This allows us to calculate the $\gamma$-ray luminosities $(L_{\gamma})$ of those sources using $\gamma$-ray photon flux and the energy spectral index $\alpha_\gamma$, following the idea as discussed in the context of ``blazar divide'' \citep{2009MNRAS.396L.105G}, given by
\begin{equation}
	L_{\gamma} = 4\pi d_{L}^{2}\frac{S_{\gamma}(\nu_{1},\nu_{2})}{(1+z)^{1-\alpha_{\gamma}}},
\end{equation}
where $z$ is the redshift, $\alpha_{\gamma}=\Gamma -1$, 
$\Gamma$ is the photon spectral index, $d_{L}$ is the luminosity distance, and $S_{\gamma}(\nu_{1},\nu_{2})$ is the $\gamma$-ray energy flux between the frequencies $\nu_{1}$ and $\nu_{2}$, calculated from the photon flux. For this computation, we adopt a $\Lambda$CDM cosmology with $h=0.7$, $\Omega_{m}=0.3$ and $\Omega_{\Lambda}=0.7$. Figure~\ref{fig:source} shows the distribution of $\Gamma$ as a function of $L_{\gamma}$ between 1 and 100 GeV. 
This $\Gamma -L_{\gamma}$ plane indicates that FSRQs are more luminous and have steeper average photon spectral index than BL Lac objects. Also, the LSP BL-Lacs have close contiguity with FSRQs in their properties of both $L_{\gamma}$ and $\Gamma$ \citep[see also][]{2012IJMPD..2150086M, 2016RAA....16...54B}. 

\begin{figure*}
	\center
	\includegraphics[width=12 cm]{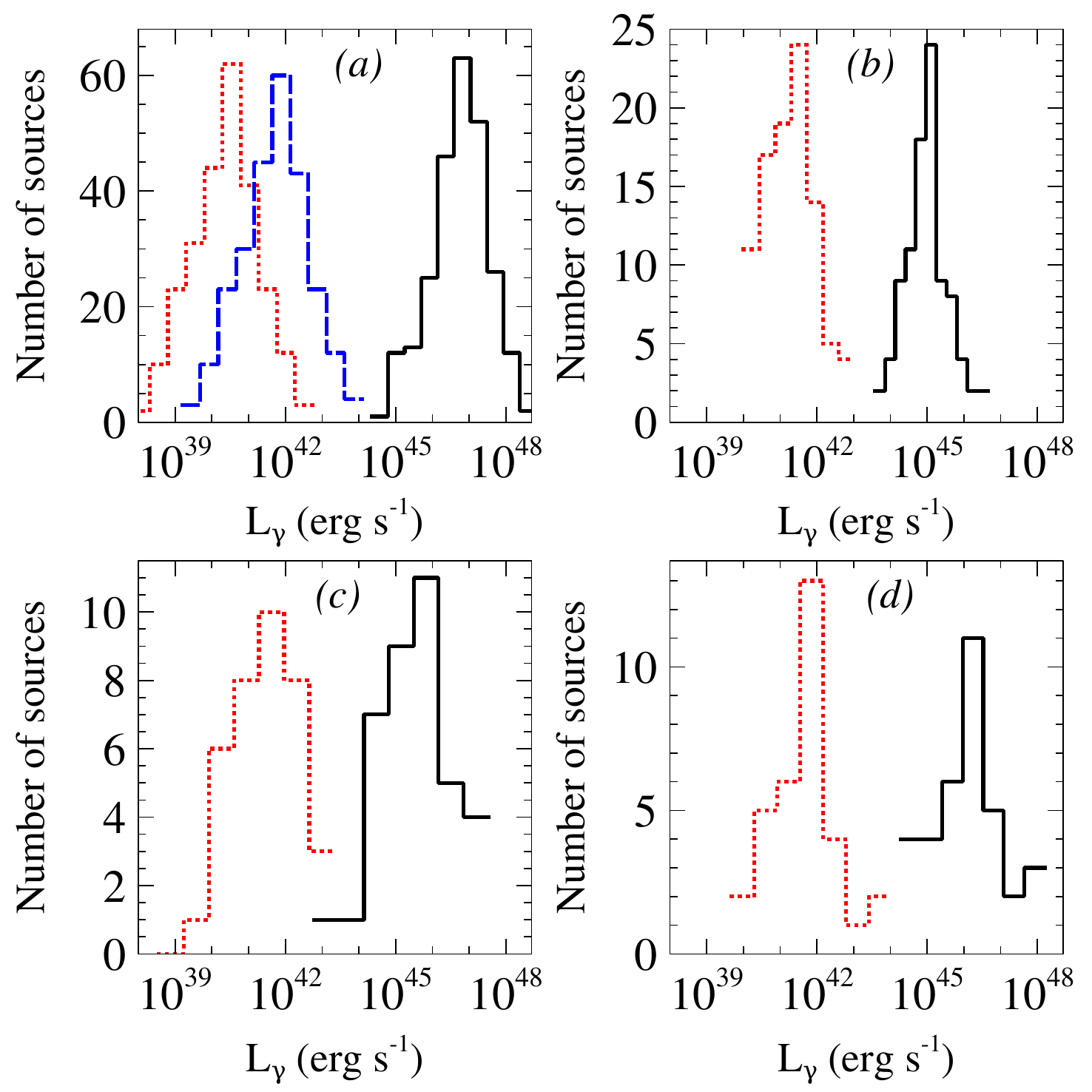}\caption{Distribution of $L_{\gamma}$ for $(a)$ FSRQs, $(b)$ HSP-BL Lacs, $(c)$ ISP-BL Lacs, and $(d)$ LSP-BL Lacs. Solid black lines are for distributions of $L_{\gamma}$ in the observer frame, while dotted or dashed lines are corresponding distributions of unbeamed or intrinsic $L_{\gamma}$. The intrinsic ones for FSRQs are computed for $95\%$ (red dotted line) and $75\%$ (blue dashed line) EC emission; those for HSP and ISP-BL Lacs are computed considering pure SSC emission; and for LSP-BL Lacs $50\%$ SSC emission is considered. Here we consider a continuous jet model with average Doppler beaming factor 20.6.} 
	\label{fig:histogram}
\end{figure*}

The blazar jets contain highly relativistic particles which are believed to be ejected either in the form of discrete blobs or as continuous flow. The observations are totally shrouded by relativistic effect, namely, beaming statistics, which is fundamentally different for two main radiation mechanisms: SSC and EC. Decomposing these beaming constraints from the intrinsic properties would allow us to probe the jet-disc symbiosis. Based on the jet ejection mechanism and beaming pattern, the observed $\gamma$-ray luminosity $L_{\text{obs}}$ and the intrinsic one $L_{\text{int}}$ are related by \citep{1995ApJ...446L..63D} 
\begin{equation}
	L_{\text{obs}}=L_{\text{int}} \delta^{m+n}, \label{eq:beam}
\end{equation}
where $m=2$ for a continuous jet and $m=3$ for a discrete jet \citep{1979ApJ...232...34B}, $\delta$ is the Doppler beaming factor, and $n$ is related to $\alpha_\gamma$ as $n=2\alpha_\gamma +1$ for EC process, whereas $n=\alpha_\gamma$ for SSC process. Here, we calculate $L_{\text{int}}$ based on different beaming constraints for all the blazar sequences. We consider a continuous jet model and an average value of $\delta=20.6$ for both FSRQs and BL Lacs \citep{2005AJ....130.1418J}. For HSP- and ISP-BL Lac objects, we consider a pure SSC emission process, while for LSP-BL Lacs  and FSRQs, a combination of SSC and EC is considered. 
However, how much EC to play role is not well constrained observationally.
Hence, we explore different plausible EC contribution in FSRQs. Figure~\ref{fig:histogram}$(a)$ distributions are for FSRQs, showing the observed $\gamma$-ray luminosity, and the intrinsic one computed for two different scenarios: $(i)$ $5\%$ SSC and $95\%$ EC, $(ii)$ $25\%$ SSC and $75\%$ EC. The distributions for HSP- and ISP-BL Lacs are described in Figure~\ref{fig:histogram}$(b)$ and $(c)$ respectively. To compute intrinsic luminosity for LSP-BL Lac objects, we consider $50\%$ SSC and $50\%$ EC emission process, as indicated in Figure~\ref{fig:histogram}$(d)$. As seen in the figure, the large difference in observed luminosities between FSRQ and BL Lac objects is significantly reduced when we compare their intrinsic luminosities. In fact, depending 
on contribution from EC, $L_{\rm int}$ for FSRQs could be even smaller 
than that of BL Lacs. This is essentially because the EC emission is much more highly beamed than the SSC emission, and also FSRQs have steeper photon index than that of BL Lac objects. Hence, while $L_{\rm obs}$ and jet power is highest
in FSRQs, their intrinsic power is significantly lower, which is in accordance 
with their Keplerian disc dominance nature.

\subsection{Explaining data by disc-outflow symbiosis} \label{sec:disc-outflow symbiosis}
We initiated the role of large scale magnetic field in a $1.5$-dimensional advective accretion flow, emphasizing that the presence of strong field could change the disc flow behaviors drastically in 
the vicinity of BH event horizon \citep{2018MNRAS.476.2396M}. Here, we consider a magnetized, viscous, 2.5-dimensional, advective disc-outflow/jet symbiotic model in a more complete framework by taking care of cooling mechanisms explicitly as discussed in Section \ref{sec:model}. As an immediate observational consequence, we apply this inflow-outflow symbiosis to classify the blazar sequences at disc-outflow/jet surface region, specifically at the jet footprint. By solving the fundamental conservation equations given in Section~\ref{sec:model},  we explore the dependence of different important cooling mechanisms on magnetic field strength and matter density. BH accretion is transonic and the relative dependence of different field components,
whose maximum possible magnitudes are constrained at sonic/critical point(s),
have been chosen in order to sustain atleast one inner saddle-type critical point and further flow dynamics.

\begin{figure*}
	\center
	\includegraphics[width=12 cm]{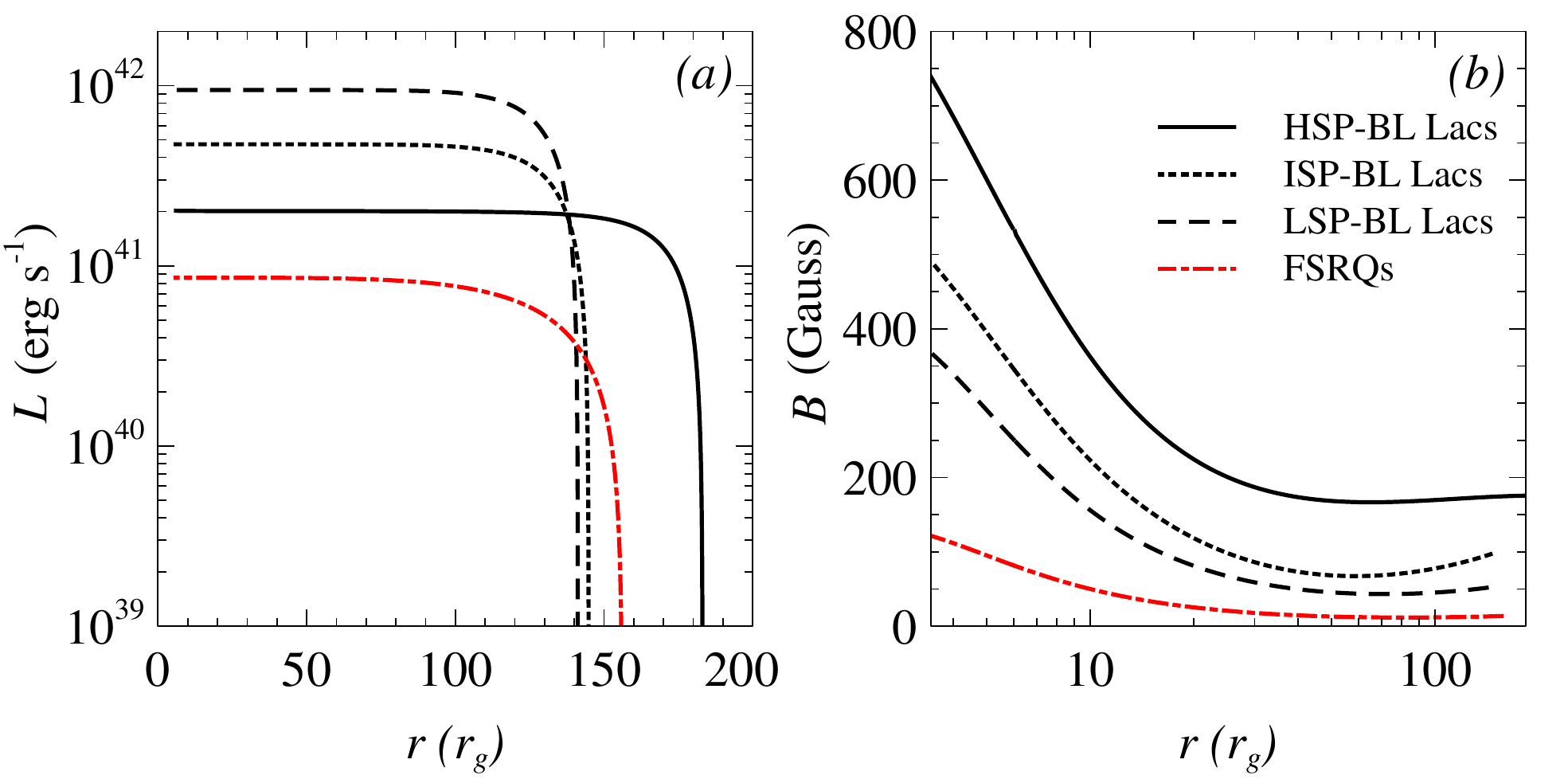}\caption{Variation of $(a)$ luminosity, and $(b)$ magnetic field strength, as functions of radial coordinate, when the different lines are for different blazar classes mentioned in $(b)$.
		The model parameters are $M=10^{8}M_{\odot}$, $\alpha = 0.01$, $\dot{m}=5\times 10^{-3}$ (FSRQs), $2\times 10^{-3}$ (LSP-BL Lacs), $10^{-3}$ (ISP-BL Lacs), $5\times 10^{-4}$ (HSP-BL Lacs), where $\dot{m}=\dot{M}/\dot{M}_{Edd}$.} 
	\label{fig:luminosity}
\end{figure*} 

\begin{figure*}
	\center
	\includegraphics[width=12 cm]{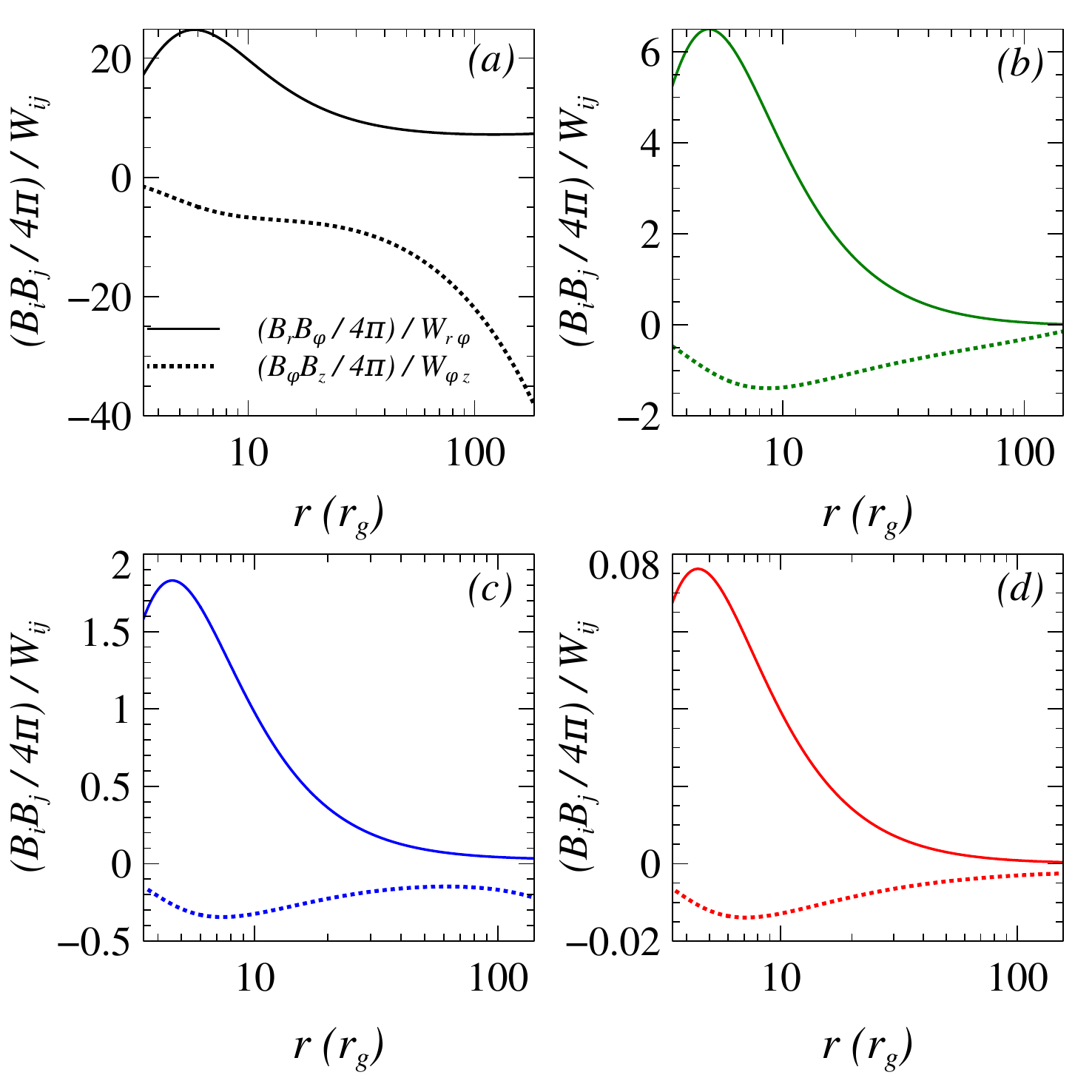}\caption{Variation of ratios of
		large scale magnetic to viscous shearing stresses for $(a)$ HSP-BL Lacs, $(b)$ ISP-BL Lacs,
		$(c)$ LSP-BL Lacs, and $(d)$ FSRQs.
		The model parameters are same as in Fig. \ref{fig:luminosity}. 
	} 
	\label{fig:shear}
\end{figure*} 

In Figure~\ref{fig:luminosity}$(a)$, we provide the luminosities obtained from all different types of 
cooling mechanisms over the entire flow, determined by various combinations of magnetic field strength and
accretion rate, whereas the corresponding magnetic field strengths are provided in Figure~\ref{fig:luminosity}$(b)$. 
The model parameters considered herein are $\alpha =0.01$, $\beta =0.3$. The vertical scalling parameters for HSP-BL Lacs are $s_{1}=-0.014$, $s_{2}=-0.0125$, $s_{3}=0.03$, $s_{4}=-0.02775$, $s_{5}=-0.021$, $s_{6}=0.01$, $s_{7}=-0.011$, and $s_{8}=-0.051$. For other classes, they are similar.
We further identify various luminosity curves therein with various blazar classes, explained as follows. Observed data indicate that the accretion rate and hence the matter density is higher in FSRQs compared to BL Lac objects. The spectra for HSP-BL Lacs lie in the extreme hard end and this hardness decreases gradually with increasing accretion rate for ISP-, LSP-BL Lacs, and finally FSRQs sequentially. 
The other important aspect in order to control blazar classes could be the magnetic field strength.  
By comparing the intrinsic luminosities of blazar classes shown in 
Fig. \ref{fig:histogram}, we argue that the matter density and magnetic field strength both play very crucial roles. For HSP-BL Lacs,
which is observed to be intrinsically lower luminous and synchrotron cooling dominated, the large magnetic field 
is expected to enhance the synchrotron cooling and also the energy amplification factor of scattered 
photon in the Comptonization process therein, which further enhances SSC mechanism. Here the magnetorotational instability (MRI, \citealt{1998RvMP...70....1B}) may be suppressed. On the other hand, for FSRQs, larger matter density helps in EC process efficiently, but their presumable 
less field reduces the synchrotron process and the underlying energy amplification factor of scattered photon, which further reduces the SSC mechanism. Hence, they are expected to be lower field systems and MRI induced $\alpha$-viscosity may operate therein. In between, for LSP-BL Lacs, the field and matter density both are expected to be optimal to operate SSC and EC mechanisms sufficiently. This makes the difference in their intrinsic luminosities. Figure \ref{fig:shear} confirms that for HSP the angular momentum
transport primarily takes place via large scale magnetic stresses, but for FSRQs it is mostly
via viscous stress govern by MRI. In between, ISP and LSP are 
expected to exhibit both the kinds of shearing stresses.




\section{DISCUSSIONS} \label{sec:discussions}

The large-scale strong magnetic field in accretion process plays very important role in angular momentum transport, radiation processes, and also in the formation of strong outflows/jets. However, the generation of such field in the accretion flow is still not well understood. 
In the accretion environment, the non-parallel gradient of temperature and density profiles may generate seed magnetic field from zero initial field condition via Biermann battery mechanism \citep{2002RvMP...74..775W}. The amplification of this seed field and the conversion between different field topologies are also possible through either differential rotation and/or turbulence mechanism and/or dynamo process \citep{1995ApJ...446..741B,1996ApJ...463..656S,2018ApJ...855..130F}. 
Also, the externally generated magnetic field from companion star or interstellar medium can be advected inward by the accreting matter and becomes dynamically dominant in the vicinity of BH through flux freezing \citep{1974Ap&SS..28...45B}.
The applications of this large scale strong field have been studied widely and also verified numerically in the `magnetically arrested disc' or MAD regime \citep{2003PASJ...55L..69N, 2011MNRAS.418L..79T}, where the large scale poloidal field plays the main role to convert mass to energy with perfect efficiency. Unlike MAD, in our differentially rotating, quasi-spherical disc-outflow symbiosis, advection of both the toroidal and poloidal field is happening. Indeed, there is an upper limit to the amount of magnetic field what the disc around a BH can sustain, given by  
$B_{Edd}\approx 10^{4} \ \text{G} \ (\frac{M}{10^{9}M_{\odot}})^{-1/2}$ \citep{2019MNRAS.482L..24M}. The required field strength to support our scenario is perfectly viable. 

The magneto-centrifugally driven outflows from the disc threaded by large scale open magnetic field is a very efficient way to extract the gravitational potential energy of BH through accretion process. Also, the positive Bernoulli number in this highly magnetized advective flow provides a generic explanation of unbound matter and, hence, strong outflows \citep{2019MNRAS.482L..24M}. Interestingly, the vertically inflated large toroidal fields create a magnetic pressure gradient, which further enhances an outward pressure to produce outflows and hence jets. Apart from disc dynamics, this large scale field also enhances the cooling mechanisms, say via synchrotron and SSC processes, very efficiently. All these magnetic natures allow us to unify the blazar classification. 

\section{CONCLUSIONS} \label{sec:conclusions}

Based on a $2.5$-dimensional geometrically thick, optically thin, magnetized, viscous, 
advective disc-outflow/jet symbiosis with cooling, we address the mutual role of matter density and large scale strong magnetic field in order to explore the energetics of the accretion induced outflow and explain blazar classes. 
We suggest that HSP-BL Lacs are more optically thin and the accretion rate and hence
matter density increases gradually for ISP-, LSP-BL Lacs, and finally FSRQs to explain their spectral signatures. To capture their intrinsic $\gamma$-ray luminosities, our model comes with the idea that the field is comparatively high in HSP-BL Lacs and this strength falls gradually for ISP- and LSP-BL Lacs, and becomes very less for FSRQs. The origin of this field and its enhancement may indicate an evolutionary unifications of blazar classes, which is beyond the scope of our model for the time being. While HSP-BL Lacs dynamics may be controlled by large scale strong magnetic fields (without MRI), FSRQs are determined by MRI.

\section*{ACKNOWLEDGEMENTS}

The authors thank Debbijoy Bhattacharya of Manipal Centre for Natural Sciences for helpful discussions on beaming corrections. The  work  was  partly  supported by the Department of Science and Technology project with Grant No. DSTO/PPH/BMP/1946 (EMR/2017/001226).




\begin{thebibliography}{}
\makeatletter
\relax
\def\mn@urlcharsother{\let\do\@makeother \do\$\do\&\do\#\do\^\do\_\do\%\do\~}
\def\mn@doi{\begingroup\mn@urlcharsother \@ifnextchar [ {\mn@doi@}
  {\mn@doi@[]}}
\def\mn@doi@[#1]#2{\def\@tempa{#1}\ifx\@tempa\@empty \href
  {http://dx.doi.org/#2} {doi:#2}\else \href {http://dx.doi.org/#2} {#1}\fi
  \endgroup}
\def\mn@eprint#1#2{\mn@eprint@#1:#2::\@nil}
\def\mn@eprint@arXiv#1{\href {http://arxiv.org/abs/#1} {{\tt arXiv:#1}}}
\def\mn@eprint@dblp#1{\href {http://dblp.uni-trier.de/rec/bibtex/#1.xml}
  {dblp:#1}}
\def\mn@eprint@#1:#2:#3:#4\@nil{\def\@tempa {#1}\def\@tempb {#2}\def\@tempc
  {#3}\ifx \@tempc \@empty \let \@tempc \@tempb \let \@tempb \@tempa \fi \ifx
  \@tempb \@empty \def\@tempb {arXiv}\fi \@ifundefined
  {mn@eprint@\@tempb}{\@tempb:\@tempc}{\expandafter \expandafter \csname
  mn@eprint@\@tempb\endcsname \expandafter{\@tempc}}}

\bibitem[\protect\citeauthoryear{{Abdo} et~al.,}{{Abdo}
  et~al.}{2010}]{2010ApJ...716...30A}
{Abdo} A.~A.,  et~al., 2010, \mn@doi [\apj] {10.1088/0004-637X/716/1/30}, \href
  {http://adsabs.harvard.edu/abs/2010ApJ...716...30A} {716, 30}

\bibitem[\protect\citeauthoryear{{Ackermann} et~al.,}{{Ackermann}
  et~al.}{2011}]{2011ApJ...743..171A}
{Ackermann} M.,  et~al., 2011, \mn@doi [\apj] {10.1088/0004-637X/743/2/171},
  \href {http://adsabs.harvard.edu/abs/2011ApJ...743..171A} {743, 171}

\bibitem[\protect\citeauthoryear{{Appl} \& {Camenzind}}{{Appl} \&
  {Camenzind}}{1992}]{1992A&A...256..354A}
{Appl} S.,  {Camenzind} M.,  1992, \aap, \href
  {http://adsabs.harvard.edu/abs/1992A%26A...256..354A} {256, 354}

\bibitem[\protect\citeauthoryear{{Balbus} \& {Hawley}}{{Balbus} \&
  {Hawley}}{1998}]{1998RvMP...70....1B}
{Balbus} S.~A.,  {Hawley} J.~F.,  1998, \mn@doi [Reviews of Modern Physics]
  {10.1103/RevModPhys.70.1}, \href
  {http://adsabs.harvard.edu/abs/1998RvMP...70....1B} {70, 1}

\bibitem[\protect\citeauthoryear{{Beskin} \& {Malyshkin}}{{Beskin} \&
  {Malyshkin}}{2000}]{2000AstL...26..208B}
{Beskin} V.~S.,  {Malyshkin} L.~M.,  2000, \mn@doi [Astronomy Letters]
  {10.1134/1.20384}, \href {http://adsabs.harvard.edu/abs/2000AstL...26..208B}
  {26, 208}

\bibitem[\protect\citeauthoryear{{Bhattacharya}, {Sreekumar}, {Mukhopadhyay}
  \& {Tomar}}{{Bhattacharya} et~al.}{2016}]{2016RAA....16...54B}
{Bhattacharya} D.,  {Sreekumar} P.,  {Mukhopadhyay} B.,   {Tomar} I.,  2016,
  \mn@doi [Research in Astronomy and Astrophysics]
  {10.1088/1674-4527/16/4/054}, \href
  {http://adsabs.harvard.edu/abs/2016RAA....16...54B} {16, 54}

\bibitem[\protect\citeauthoryear{{Bisnovatyi-Kogan} \&
  {Ruzmaikin}}{{Bisnovatyi-Kogan} \& {Ruzmaikin}}{1974}]{1974Ap&SS..28...45B}
{Bisnovatyi-Kogan} G.~S.,  {Ruzmaikin} A.~A.,  1974, \mn@doi [\apss]
  {10.1007/BF00642237}, \href
  {http://adsabs.harvard.edu/abs/1974Ap%26SS..28...45B} {28, 45}

\bibitem[\protect\citeauthoryear{{Blandford} \& {K{\"o}nigl}}{{Blandford} \&
  {K{\"o}nigl}}{1979}]{1979ApJ...232...34B}
{Blandford} R.~D.,  {K{\"o}nigl} A.,  1979, \mn@doi [\apj] {10.1086/157262},
  \href {http://adsabs.harvard.edu/abs/1979ApJ...232...34B} {232, 34}

\bibitem[\protect\citeauthoryear{{Blandford} \& {Payne}}{{Blandford} \&
  {Payne}}{1982}]{1982MNRAS.199..883B}
{Blandford} R.~D.,  {Payne} D.~G.,  1982, \mn@doi [\mnras]
  {10.1093/mnras/199.4.883}, \href
  {http://adsabs.harvard.edu/abs/1982MNRAS.199..883B} {199, 883}

\bibitem[\protect\citeauthoryear{{Blandford} \& {Znajek}}{{Blandford} \&
  {Znajek}}{1977}]{1977MNRAS.179..433B}
{Blandford} R.~D.,  {Znajek} R.~L.,  1977, \mn@doi [\mnras]
  {10.1093/mnras/179.3.433}, \href
  {http://adsabs.harvard.edu/abs/1977MNRAS.179..433B} {179, 433}

\bibitem[\protect\citeauthoryear{{B{\l}a{\.z}ejowski}, {Sikora}, {Moderski}  \&
  {Madejski}}{{B{\l}a{\.z}ejowski} et~al.}{2000}]{2000ApJ...545..107B}
{B{\l}a{\.z}ejowski} M.,  {Sikora} M.,  {Moderski} R.,   {Madejski} G.~M.,
  2000, \mn@doi [\apj] {10.1086/317791}, \href
  {http://adsabs.harvard.edu/abs/2000ApJ...545..107B} {545, 107}

\bibitem[\protect\citeauthoryear{{Brandenburg}, {Nordlund}, {Stein}  \&
  {Torkelsson}}{{Brandenburg} et~al.}{1995}]{1995ApJ...446..741B}
{Brandenburg} A.,  {Nordlund} A.,  {Stein} R.~F.,   {Torkelsson} U.,  1995,
  \mn@doi [\apj] {10.1086/175831}, \href
  {http://adsabs.harvard.edu/abs/1995ApJ...446..741B} {446, 741}

\bibitem[\protect\citeauthoryear{{Dermer}}{{Dermer}}{1995}]{1995ApJ...446L..63D}
{Dermer} C.~D.,  1995, \mn@doi [\apjl] {10.1086/187931}, \href
  {http://adsabs.harvard.edu/abs/1995ApJ...446L..63D} {446, L63}

\bibitem[\protect\citeauthoryear{{Dermer} \& {Schlickeiser}}{{Dermer} \&
  {Schlickeiser}}{1993}]{1993ApJ...416..458D}
{Dermer} C.~D.,  {Schlickeiser} R.,  1993, \mn@doi [\apj] {10.1086/173251},
  \href {http://adsabs.harvard.edu/abs/1993ApJ...416..458D} {416, 458}

\bibitem[\protect\citeauthoryear{{Dermer}, {Liang}  \& {Canfield}}{{Dermer}
  et~al.}{1991}]{1991ApJ...369..410D}
{Dermer} C.~D.,  {Liang} E.~P.,   {Canfield} E.,  1991, \mn@doi [\apj]
  {10.1086/169770}, \href {http://adsabs.harvard.edu/abs/1991ApJ...369..410D}
  {369, 410}

\bibitem[\protect\citeauthoryear{{Eatough} et~al.,}{{Eatough}
  et~al.}{2013}]{2013Natur.501..391E}
{Eatough} R.~P.,  et~al., 2013, \mn@doi [\nat] {10.1038/nature12499}, \href
  {http://adsabs.harvard.edu/abs/2013Natur.501..391E} {501, 391}

\bibitem[\protect\citeauthoryear{{Fender}, {Belloni}  \& {Gallo}}{{Fender}
  et~al.}{2004}]{2004MNRAS.355.1105F}
{Fender} R.~P.,  {Belloni} T.~M.,   {Gallo} E.,  2004, \mn@doi [\mnras]
  {10.1111/j.1365-2966.2004.08384.x}, \href
  {http://adsabs.harvard.edu/abs/2004MNRAS.355.1105F} {355, 1105}

\bibitem[\protect\citeauthoryear{{Fendt} \& {Ga{\ss}mann}}{{Fendt} \&
  {Ga{\ss}mann}}{2018}]{2018ApJ...855..130F}
{Fendt} C.,  {Ga{\ss}mann} D.,  2018, \mn@doi [\apj]
  {10.3847/1538-4357/aab14c}, \href
  {http://adsabs.harvard.edu/abs/2018ApJ...855..130F} {855, 130}

\bibitem[\protect\citeauthoryear{{Ghisellini}, {Maraschi}  \&
  {Tavecchio}}{{Ghisellini} et~al.}{2009}]{2009MNRAS.396L.105G}
{Ghisellini} G.,  {Maraschi} L.,   {Tavecchio} F.,  2009, \mn@doi [\mnras]
  {10.1111/j.1745-3933.2009.00673.x}, \href
  {http://adsabs.harvard.edu/abs/2009MNRAS.396L.105G} {396, L105}

\bibitem[\protect\citeauthoryear{{Ghisellini}, {Tavecchio}, {Maraschi},
  {Celotti}  \& {Sbarrato}}{{Ghisellini} et~al.}{2014}]{2014Natur.515..376G}
{Ghisellini} G.,  {Tavecchio} F.,  {Maraschi} L.,  {Celotti} A.,   {Sbarrato}
  T.,  2014, \mn@doi [\nat] {10.1038/nature13856}, \href
  {http://adsabs.harvard.edu/abs/2014Natur.515..376G} {515, 376}

\bibitem[\protect\citeauthoryear{{Guilet} \& {Ogilvie}}{{Guilet} \&
  {Ogilvie}}{2012}]{2012MNRAS.424.2097G}
{Guilet} J.,  {Ogilvie} G.~I.,  2012, \mn@doi [\mnras]
  {10.1111/j.1365-2966.2012.21361.x}, \href
  {http://adsabs.harvard.edu/abs/2012MNRAS.424.2097G} {424, 2097}

\bibitem[\protect\citeauthoryear{{Jorstad} et~al.,}{{Jorstad}
  et~al.}{2005}]{2005AJ....130.1418J}
{Jorstad} S.~G.,  et~al., 2005, \mn@doi [\apj] {10.1086/444593}, \href
  {http://adsabs.harvard.edu/abs/2005AJ....130.1418J} {130, 1418}

\bibitem[\protect\citeauthoryear{{Kuncic} \& {Bicknell}}{{Kuncic} \&
  {Bicknell}}{2004}]{2004ApJ...616..669K}
{Kuncic} Z.,  {Bicknell} G.~V.,  2004, \mn@doi [\apj] {10.1086/425032}, \href
  {http://adsabs.harvard.edu/abs/2004ApJ...616..669K} {616, 669}

\bibitem[\protect\citeauthoryear{{Lynden-Bell}}{{Lynden-Bell}}{2003}]{2003MNRAS.341.1360L}
{Lynden-Bell} D.,  2003, \mn@doi [\mnras] {10.1046/j.1365-8711.2003.06506.x},
  \href {http://adsabs.harvard.edu/abs/2003MNRAS.341.1360L} {341, 1360}

\bibitem[\protect\citeauthoryear{{Ma}, {Xie}  \& {Hou}}{{Ma}
  et~al.}{2014}]{2014ApJ...780L..14M}
{Ma} R.,  {Xie} F.-G.,   {Hou} S.,  2014, \mn@doi [\apjl]
  {10.1088/2041-8205/780/1/L14}, \href
  {http://adsabs.harvard.edu/abs/2014ApJ...780L..14M} {780, L14}

\bibitem[\protect\citeauthoryear{{McKinney} \& {Gammie}}{{McKinney} \&
  {Gammie}}{2004}]{2004ApJ...611..977M}
{McKinney} J.~C.,  {Gammie} C.~F.,  2004, \mn@doi [\apj] {10.1086/422244},
  \href {http://adsabs.harvard.edu/abs/2004ApJ...611..977M} {611, 977}

\bibitem[\protect\citeauthoryear{{Meier}, {Koide}  \& {Uchida}}{{Meier}
  et~al.}{2001}]{2001Sci...291...84M}
{Meier} D.~L.,  {Koide} S.,   {Uchida} Y.,  2001, \mn@doi [Science]
  {10.1126/science.291.5501.84}, \href
  {http://adsabs.harvard.edu/abs/2001Sci...291...84M} {291, 84}

\bibitem[\protect\citeauthoryear{{Mondal} \& {Mukhopadhyay}}{{Mondal} \&
  {Mukhopadhyay}}{2018}]{2018MNRAS.476.2396M}
{Mondal} T.,  {Mukhopadhyay} B.,  2018, \mn@doi [\mnras]
  {10.1093/mnras/sty332}, \href
  {http://adsabs.harvard.edu/abs/2018MNRAS.476.2396M} {476, 2396}

\bibitem[\protect\citeauthoryear{{Mondal} \& {Mukhopadhyay}}{{Mondal} \&
  {Mukhopadhyay}}{2019}]{2019MNRAS.482L..24M}
{Mondal} T.,  {Mukhopadhyay} B.,  2019, \mn@doi [\mnras]
  {10.1093/mnrasl/sly165}, \href
  {http://adsabs.harvard.edu/abs/2019MNRAS.482L..24M} {482, L24}

\bibitem[\protect\citeauthoryear{{Mukhopadhyay}}{{Mukhopadhyay}}{2002}]{2002ApJ...581..427M}
{Mukhopadhyay} B.,  2002, \mn@doi [\apj] {10.1086/344227}, \href
  {http://adsabs.harvard.edu/abs/2002ApJ...581..427M} {581, 427}

\bibitem[\protect\citeauthoryear{{Mukhopadhyay}, {Bhattacharya}  \&
  {Sreekumar}}{{Mukhopadhyay} et~al.}{2012}]{2012IJMPD..2150086M}
{Mukhopadhyay} B.,  {Bhattacharya} D.,   {Sreekumar} P.,  2012, \mn@doi
  [International Journal of Modern Physics D] {10.1142/S0218271812500861},
  \href {http://adsabs.harvard.edu/abs/2012IJMPD..2150086M} {21, 1250086}

\bibitem[\protect\citeauthoryear{{Nagar}, {Falcke}, {Wilson}  \& {Ho}}{{Nagar}
  et~al.}{2000}]{2000ApJ...542..186N}
{Nagar} N.~M.,  {Falcke} H.,  {Wilson} A.~S.,   {Ho} L.~C.,  2000, \mn@doi
  [\apj] {10.1086/309524}, \href
  {http://adsabs.harvard.edu/abs/2000ApJ...542..186N} {542, 186}

\bibitem[\protect\citeauthoryear{{Narayan} \& {Yi}}{{Narayan} \&
  {Yi}}{1995}]{1995ApJ...452..710N}
{Narayan} R.,  {Yi} I.,  1995, \mn@doi [\apj] {10.1086/176343}, \href
  {http://adsabs.harvard.edu/abs/1995ApJ...452..710N} {452, 710}

\bibitem[\protect\citeauthoryear{{Narayan}, {Igumenshchev}  \&
  {Abramowicz}}{{Narayan} et~al.}{2003}]{2003PASJ...55L..69N}
{Narayan} R.,  {Igumenshchev} I.~V.,   {Abramowicz} M.~A.,  2003, \mn@doi
  [PASJ] {10.1093/pasj/55.6.L69}, \href
  {http://adsabs.harvard.edu/abs/2003PASJ...55L..69N} {55, L69}

\bibitem[\protect\citeauthoryear{{Nemmen}, {Georganopoulos}, {Guiriec},
  {Meyer}, {Gehrels}  \& {Sambruna}}{{Nemmen}
  et~al.}{2012}]{2012Sci...338.1445N}
{Nemmen} R.~S.,  {Georganopoulos} M.,  {Guiriec} S.,  {Meyer} E.~T.,  {Gehrels}
  N.,   {Sambruna} R.~M.,  2012, \mn@doi [Science] {10.1126/science.1227416},
  \href {http://adsabs.harvard.edu/abs/2012Sci...338.1445N} {338, 1445}

\bibitem[\protect\citeauthoryear{{Novikov} \& {Thorne}}{{Novikov} \&
  {Thorne}}{1973}]{1973blho.conf..343N}
{Novikov} I.~D.,  {Thorne} K.~S.,  1973, in {Dewitt} C.,  {Dewitt} B.~S.,  eds,
  Black Holes (Les Astres Occlus). pp 343--450

\bibitem[\protect\citeauthoryear{{Qian}, {Fendt}  \& {Vourellis}}{{Qian}
  et~al.}{2018}]{2018ApJ...859...28Q}
{Qian} Q.,  {Fendt} C.,   {Vourellis} C.,  2018, \mn@doi [\apj]
  {10.3847/1538-4357/aabd36}, \href
  {http://adsabs.harvard.edu/abs/2018ApJ...859...28Q} {859, 28}

\bibitem[\protect\citeauthoryear{{Rajesh} \& {Mukhopadhyay}}{{Rajesh} \&
  {Mukhopadhyay}}{2010}]{2010MNRAS.402..961R}
{Rajesh} S.~R.,  {Mukhopadhyay} B.,  2010, \mn@doi [\mnras]
  {10.1111/j.1365-2966.2009.15925.x}, \href
  {http://adsabs.harvard.edu/abs/2010MNRAS.402..961R} {402, 961}

\bibitem[\protect\citeauthoryear{{Shakura} \& {Sunyaev}}{{Shakura} \&
  {Sunyaev}}{1973}]{1973A&A....24..337S}
{Shakura} N.~I.,  {Sunyaev} R.~A.,  1973, \aap, \href
  {http://adsabs.harvard.edu/abs/1973A%26A....24..337S} {24, 337}

\bibitem[\protect\citeauthoryear{{Sikora}, {Stawarz}, {Moderski}, {Nalewajko}
  \& {Madejski}}{{Sikora} et~al.}{2009}]{2009ApJ...704...38S}
{Sikora} M.,  {Stawarz} {\L}.,  {Moderski} R.,  {Nalewajko} K.,   {Madejski}
  G.~M.,  2009, \mn@doi [\apj] {10.1088/0004-637X/704/1/38}, \href
  {http://adsabs.harvard.edu/abs/2009ApJ...704...38S} {704, 38}

\bibitem[\protect\citeauthoryear{{Stone}, {Hawley}, {Gammie}  \&
  {Balbus}}{{Stone} et~al.}{1996}]{1996ApJ...463..656S}
{Stone} J.~M.,  {Hawley} J.~F.,  {Gammie} C.~F.,   {Balbus} S.~A.,  1996,
  \mn@doi [\apj] {10.1086/177280}, \href
  {http://adsabs.harvard.edu/abs/1996ApJ...463..656S} {463, 656}

\bibitem[\protect\citeauthoryear{{Tchekhovskoy}, {Narayan}  \&
  {McKinney}}{{Tchekhovskoy} et~al.}{2011}]{2011MNRAS.418L..79T}
{Tchekhovskoy} A.,  {Narayan} R.,   {McKinney} J.~C.,  2011, \mn@doi [\mnras]
  {10.1111/j.1745-3933.2011.01147.x}, \href
  {http://adsabs.harvard.edu/abs/2011MNRAS.418L..79T} {418, L79}

\bibitem[\protect\citeauthoryear{{Urry} \& {Padovani}}{{Urry} \&
  {Padovani}}{1995}]{1995PASP..107..803U}
{Urry} C.~M.,  {Padovani} P.,  1995, \mn@doi [PASP] {10.1086/133630}, \href
  {http://adsabs.harvard.edu/abs/1995PASP..107..803U} {107, 803}

\bibitem[\protect\citeauthoryear{{Widrow}}{{Widrow}}{2002}]{2002RvMP...74..775W}
{Widrow} L.~M.,  2002, \mn@doi [Reviews of Modern Physics]
  {10.1103/RevModPhys.74.775}, \href
  {http://adsabs.harvard.edu/abs/2002RvMP...74..775W} {74, 775}

\bibitem[\protect\citeauthoryear{{Zamaninasab}, {Clausen-Brown}, {Savolainen}
  \& {Tchekhovskoy}}{{Zamaninasab} et~al.}{2014}]{2014Natur.510..126Z}
{Zamaninasab} M.,  {Clausen-Brown} E.,  {Savolainen} T.,   {Tchekhovskoy} A.,
  2014, \mn@doi [\nat] {10.1038/nature13399}, \href
  {http://adsabs.harvard.edu/abs/2014Natur.510..126Z} {510, 126}

\makeatother
\end{thebibliography}







\bsp	
\label{lastpage}
\end{document}